# RSA SIGNATURE: BEHIND THE SCENES

Dragan Vidakovic[1], Dusko Parezanovic[1], Olivera Nikolic[2] and Jelena Kaljevic[2]

[1]Gimnazija Ivanjica, Serbia
[2] Faculty of Business Valjevo, Singidunum University Belgrade, Serbia
dragan.vidakovic@open.telekom.rs
infomat@open.telekom.rs
{onikolic,jkaljevic}@singidunum.ac.rs

## *ABSTRACT*

*In this paper, we present a complete digital signature message stream, just the way the RSA digital signature scheme does it. We will focus on the operations with large numbers due to the fact that operating with large numbers is the essence of RSA that cannot be understood by the usual illustrative examples with small numbers[1].*

## 1. INTRODUCTION

The idea of RSA is based on the belief that it is difficult to factor the number that is the product of two large prime numbers. Because of that it is necessary to develop the arithmetic of large numbers operations, as well as to encode the algorithm for number primality test, a hash function and many more auxiliary functions that are necessary for developing of the own digital signature software[4].

Many people have heard about a digital signature and read a notice saying that a document is digitally signed, but few of them have a real idea of what a digital signature is and how it looks like.

Below, we will present in detail how to generate a digital signature. We are sure that this will be an inspiring step for many people to try to develop their own tools for the protection of their data integrity.

## 2. THE RSA SIGNATURE SCHEME

In this paragraph, we will recall the steps that are necessary for the RSA scheme [2][3].

**Algorithm** Key generation for the RSA signature scheme
SUMMARY: each entity creates an RSA public key and a corresponding private key.
Each entity A should do the following:
**1.** Generate two large distinct random primes p and q, each roughly the same size (see x11.3.2).
2. Compute $n = pq$ and $\phi = (p - 1)(q - 1)$.
3. Select a random integer e, $1 < e < \phi$ such that $\gcd(e, \phi) = 1$.
4. Use the extended Euclidean algorithm ([2]) to compute the unique integer d, $1 < d < \phi$, such that $ed \equiv 1 \pmod{\phi}$
5. A’s public key is (n; e); A’s private key is d

**Algorithm** RSA signature generation and verification
SUMMARY: entity A signs a message $m \in \mathbf{W}$. Any entity B can verify A’s signature and recover the message m from the signature.
1. *Signature generation.* Entity A should do the following:
(a) Compute **m'** = R(m), an integer in the range [0; n − 1].
(b) Compute **s** = $(\mathbf{m'})^{\mathbf{d}}$ mod **n**.
(c) A's signature for m is s.

2. *Verification.* To verify A's signature s and recover the message m, B should:
(a) Obtain A's authentic public key (n; e).
(b) Compute **m'**= $\mathbf{s}^{\mathbf{e}}$ mod **n**.
(c) Verify that $\mathbf{m'} \in \mathbf{W}_{\mathbf{R}}$; if not, reject the signature.
(d) Recover m = R−1(**m'**).

## 3. PREPARATORY STEP

In order to sign a message, we need to prepare many functions. Since Hash value of the message is central in the digital signature, we consider it is very important that we have a software for finding hash value.

In this paragraph, we will show the algotithm and code for SHA-1.

### 3.1. SECURE HASH ALGORITHM (SHA-1)

In this paragraph we specify SHA-1 [2], for several reasons: Because of the digital signature, to see how seem complicated and daunting and in the end because we can see how it can be solved by simply tools such as Delphi 7 console application.

INPUT: bitstring x of bitlength $b \geq 0$.
OUTPUT: 160-bit hash-code of x.
1. Definition of constans. Define a fifth (32-bit initial chaining values) IV to match those in MD4: h5 = 0xc3d2e1f0. h5 = 0xc3d2e1f0.
Define per-round integer additive constants: y1 = 0x5a827999, y2 = 0x6ed9eba1,
y3 = 0x8f1bbcdc, y4 = 0xca62c1d6. (No order for accessing source words, or specification of bit positions for left shifts is required.)

2. Overall preprocessing. Pad as in MD4, except the final two 32-bit words specifying the bitlength b is appended with most significant word preceding least significant.
As in MD4, the formatted input is 16m 32-bit words: $x_0 x_1 \ldots x_{16m-1}$. Initialize chaining variables: $(H_1;H_2;H_3;H_4;H_5) \leftarrow (h_1; h_2; h_3; h_4; h_5)$.

3. Processing. For each i from 0 to m − 1, copy the $i^{th}$ block of sixteen 32-bit words into temporary storage: $X[j] \leftarrow x_{16i+j}$; $0 \leq j \leq 15$, and process these as below in four 20-step rounds before updating the chaining variables
(expand 16-word block into 80-word block; let Xj denote X[j])
for j from 16 to 79, $X_j \leftarrow ((X_{j-3}(X_{j-8} \oplus X_{j-14} \oplus X_{j-16}) \hookleftarrow 1)$.
(initialize working variables) $(A, B, C, D, E) \leftarrow (H_1, H_2, H_3, H_4, H_5)$.

(Round 1) For j from 0 to 19 do the following:
$t \leftarrow ((A \hookleftarrow 5) + f(B, C, D) + E + X_j + y_1)$,
$(A, B, C, D, E) \leftarrow (t, A, B \hookleftarrow 30, C, D)$.

(Round 2) For j from 20 to 39 do the following
$t \leftarrow ((A \hookleftarrow 5) + h(B, C, D) + E + X_j + y_2)$.
$(A, B, C, D, E) \leftarrow (t, A, B \hookleftarrow 30, C, D)$.

(Round 3) For j from 40 to 59 do the following:

$t \leftarrow ((A \hookleftarrow 5) + g(B,C,D) + E + X_j + y_3)$
$(A, B, C, D, E) \leftarrow (t, A, B \hookleftarrow 30, C, D)$.

(Round 4) For j from 60 to 79 do the following:
$t \leftarrow ((A \hookleftarrow 5) + h(B, C, D) + E + X_j + y_4)$.
$(A, B, C, D, E) \leftarrow (t, A, B \hookleftarrow 30, C, D)$.
(update chaining values)
$(H_1, H_2, H_3, H_4, H_5) \leftarrow (H_1 + A, H_2 + B, H_3 + C, H_4 + D, H_5 + E)$.

4. Completion. The hash-value is: $H_1 \,\&\, H_2 \,\&\, H_3 \,\&\, H_4 \,\&\, H_5$.
(with first and last bytes the high- and low-order bytes of H1, H5, respectively)

Where:
& : concatenation of strings
**+ :** addition modulo $2^{32}$
$f(u,v,w) = uv \vee u'w$
$g(u,v,w) = uv \vee uw \vee vw$
$h(u,v,w) = u \oplus v \oplus w$
uv: and
u' : complement
$u \vee v$ : or
$\oplus$**:** exclusive or
$u \hookleftarrow s$ **:** rotation to the left for s position
$(X_1, \ldots, X_j) \leftarrow (Y_1, \ldots, Y_j)$ : simultaneous assignment $(X_i \leftarrow Y_i)$.

## 3.2. CODE FOR SHA-1

In this paragraph, we will encode upper algorithm. We will use console application Delphi 7.

```
PROGRAM SHA_1;
{$APPTYPE CONSOLE}
var c1: char;
k,i,j,l,duz,duz1,m,I1,I2,I3,I4:integer;
a:array[1..8] of integer;
a1,a2:array[1..32] of integer;
h1,h2,h3,h4,h5,y1,y2,y3,y4,hh1,hh2,hh3,hh4,hh5,p:array [0..31] of integer;
aa,bb,cc,dd,ee,pp,qq,rr,tt,ss,nn,mm:array[0..31] of integer;
pom:array[0..35] of integer;
x:array[0..79,0..31] of integer;
f,g:file of integer;

procedure dodeli(var a:array of integer;b:array of integer);
var i:integer;
begin
for i:=0 to 31 do a[i]:=b[i];
end;
procedure rot(var a:array of integer;t:integer);
var i,k,l:integer;
begin
for i:=1 to t do
begin
k:=a[0];
for l:=0 to 30 do a[l]:=a[l+1];
```

```
a[31]:=k;
end;
end;
procedure kom(var a:array of integer);
var i,j:integer;
begin
for i:=0 to 31 do
if a[i]=0 then a[i]:=1
else a[i]:=0;
end;
procedure fi(u,v,w:array of integer;var t:array of integer);
var i,j:integer;
p:array[0..31] of integer;
begin
for i:=0 to 31 do v[i]:=v[i] and u[i];
kom(u);
for i:=0 to 31 do t[i]:=v[i] or (u[i] and w[i]);
end;
procedure gi(u,v,w:array of integer;var t:array of integer);
var i,j:integer;
begin
for i:=0 to 31 do t[i]:=(u[i] and v[i]) or (u[i] and w[i]) or (v[i] and w[i]);
end;
procedure hi(u,v,w:array of integer;var t:array of integer);
var i,j:integer;
begin
for i:=0 to 31 do t[i]:=(u[i] xor v[i]) xor w[i];
end;
procedure saberi(a,b:array of integer;var w:array of integer);
var c:integer;
begin
c:=0;
for i:=31 downto 0 do
begin
w[i]:=(a[i]+b[i]+c) mod 2;
if (a[i]+b[i]+c)<2 then c:=0
else c:=1;
end;
end;
procedure ses(a,b,c,d:integer);
var s:integer;
begin
s:=0;
s:=a*8+b*4+c*2+d;
if s=0 then write('0');if s=1 then write('1');if s=2 then write('2');
if s=3 then write('3');if s=4 then write('4');if s=5 then write('5');
if s=6 then write('6');if s=7 then write('7');if s=8 then write('8');
if s=9 then write('9');if s=10 then write('a');if s=11 then write('b');
if s=12 then write('c');if s=13 then write('d');if s=14 then write('e');
if s=15 then write('f');
end;
```

```
begin
 writeln;
writeln('Type your message to 147 symbols- because we use EOLN-Enter. For larger messages we can
use files');
assign(g,'por.dat');
rewrite(g);
duz:=0;
writeln;
write('Input message:');
while not eoln do
begin
    read(c1);
    k:=ord(c1);
    for i:=1 to 8 do a[i]:=0;
    i:=1;
   while k<>0 do
    begin
     a[i]:=k mod 2;
     k:=k div 2;
     i:=i+1;
    end;
 duz:=duz+8;
  for I:=8 downto 1 do write(g,A[I]);
end;
{Padding}
  duz1:=duz;
  k:=1;
  l:=0;
  write(g,k);
  duz:=duz+1;
if duz mod 512=0 then
begin
 for i:=1 to 512-64 do write(g,l);
 duz:=duz+512-64;
 end
 else
 begin
  k:=duz mod 512;
  for i:=1 to 512-k-64 do write(g,l);
  duz:=duz+512-k-64;
  end;
 i:=1;
 while duz1<>0 do
 begin
  if i<=32 then
   begin
    a1[i]:=duz1 mod 2;
    duz1:=duz1 div 2
    end
    else
```

```
    begin
    a2[i]:=duz1 mod 2;
    duz1:=duz1 div 2;
    end;
    i:=i+1;
  end;
   for i:=32 downto 1 do write(g,a2[i]);
   for i:=32 downto 1 do write(g,a1[i]);
  {big-endian }
    {end of pading}
     {Defining Constants}
         { Constants do not have to recalculate}
 h1[31]:=1;h1[30]:=0;h1[29]:=0;h1[28]:=0; h1[27]:=0;h1[26]:=0;h1[25]:=0;h1[24]:=0;
h1[23]:=1;h1[22]:=1;h1[21]:=0;h1[20]:=0; h1[19]:=0;h1[18]:=1;h1[17]:=0;h1[16]:=0;
h1[15]:=1;h1[14]:=0;h1[13]:=1;h1[12]:=0; h1[11]:=0;h1[10]:=0;h1[9]:=1;h1[8]:=0;
h1[7]:=1;h1[6]:=1;h1[5]:=1;h1[4]:=0; h1[3]:=0;h1[2]:=1;h1[1]:=1;h1[0]:=0;

h2[31]:=1;h2[30]:=0;h2[29]:=0;h2[28]:=1; h2[27]:=0;h2[26]:=0;h2[25]:=0;h2[24]:=1;
h2[23]:=1;h2[22]:=1;h2[21]:=0;h2[20]:=1; h2[19]:=0;h2[18]:=1;h2[17]:=0;h2[16]:=1;
h2[15]:=1;h2[14]:=0;h2[13]:=1;h2[12]:=1; h2[11]:=0;h2[10]:=0;h2[9]:=1;h2[8]:=1;
h2[7]:=1;h2[6]:=1;h2[5]:=1;h2[4]:=1; h2[3]:=0;h2[2]:=1;h2[1]:=1;h2[0]:=1;

h3[31]:=0;h3[30]:=1;h3[29]:=1;h3[28]:=1; h3[27]:=1;h3[26]:=1;h3[25]:=1;h3[24]:=1;
h3[23]:=0;h3[22]:=0;h3[21]:=1;h3[20]:=1; h3[19]:=1;h3[18]:=0;h3[17]:=1;h3[16]:=1;
h3[15]:=0;h3[14]:=1;h3[13]:=0;h3[12]:=1; h3[11]:=1;h3[10]:=1;h3[9]:=0;h3[8]:=1;
h3[7]:=0;h3[6]:=0;h3[5]:=0;h3[4]:=1; h3[3]:=1;h3[2]:=0;h3[1]:=0;h3[0]:=1;

h4[31]:=0;h4[30]:=1;h4[29]:=1;h4[28]:=0; h4[27]:=1;h4[26]:=1;h4[25]:=1;h4[24]:=0;
h4[23]:=0;h4[22]:=0;h4[21]:=1;h4[20]:=0; h4[19]:=1;h4[18]:=0;h4[17]:=1;h4[16]:=0;
h4[15]:=0;h4[14]:=1;h4[13]:=0;h4[12]:=0; h4[11]:=1;h4[10]:=1;h4[9]:=0;h4[8]:=0;
h4[7]:=0;h4[6]:=0;h4[5]:=0;h4[4]:=0; h4[3]:=1;h4[2]:=0;h4[1]:=0;h4[0]:=0;

h5[31]:=0;h5[30]:=0;h5[29]:=0;h5[28]:=0; h5[27]:=1;h5[26]:=1;h5[25]:=1;h5[24]:=1;
h5[23]:=1;h5[22]:=0;h5[21]:=0;h5[20]:=0; h5[19]:=0;h5[18]:=1;h5[17]:=1;h5[16]:=1;
h5[15]:=0;h5[14]:=1;h5[13]:=0;h5[12]:=0; h5[11]:=1;h5[10]:=0;h5[9]:=1;h5[8]:=1;
h5[7]:=1;h5[6]:=1;h5[5]:=0;h5[4]:=0; h5[3]:=0;h5[2]:=0;h5[1]:=1;h5[0]:=1;

y1[31]:=1;y1[30]:=0;y1[29]:=0;y1[28]:=1; y1[27]:=1;y1[26]:=0;y1[25]:=0;y1[24]:=1;
y1[23]:=1;y1[22]:=0;y1[21]:=0;y1[20]:=1; y1[19]:=1;y1[18]:=1;y1[17]:=1;y1[16]:=0;
y1[15]:=0;y1[14]:=1;y1[13]:=0;y1[12]:=0; y1[11]:=0;y1[10]:=0;y1[9]:=0;y1[8]:=1;
y1[7]:=0;y1[6]:=1;y1[5]:=0;y1[4]:=1; y1[3]:=1;y1[2]:=0;y1[1]:=1;y1[0]:=0;

y2[31]:=1;y2[30]:=0;y2[29]:=0;y2[28]:=0; y2[27]:=0;y2[26]:=1;y2[25]:=0;y2[24]:=1;
y2[23]:=1;y2[22]:=1;y2[21]:=0;y2[20]:=1; y2[19]:=0;y2[18]:=1;y2[17]:=1;y2[16]:=1;
y2[15]:=1;y2[14]:=0;y2[13]:=0;y2[12]:=1; y2[11]:=1;y2[10]:=0;y2[9]:=1;y2[8]:=1;
y2[7]:=0;y2[6]:=1;y2[5]:=1;y2[4]:=1; y2[3]:=0;y2[2]:=1;y2[1]:=1;y2[0]:=0;

y3[31]:=0;y3[30]:=0;y3[29]:=1;y3[28]:=1; y3[27]:=1;y3[26]:=0;y3[25]:=1;y3[24]:=1;
y3[23]:=0;y3[22]:=0;y3[21]:=1;y3[20]:=1; y3[19]:=1;y3[18]:=1;y3[17]:=0;y3[16]:=1;
y3[15]:=1;y3[14]:=1;y3[13]:=0;y3[12]:=1; y3[11]:=1;y3[10]:=0;y3[9]:=0;y3[8]:=0;
y3[7]:=1;y3[6]:=1;y3[5]:=1;y3[4]:=1; y3[3]:=0;y3[2]:=0;y3[1]:=0;y3[0]:=1;
```

```
y4[31]:=0;y4[30]:=1;y4[29]:=1;y4[28]:=0; y4[27]:=1;y4[26]:=0;y4[25]:=1;y4[24]:=1;
y4[23]:=1;y4[22]:=0;y4[21]:=0;y4[20]:=0; y4[19]:=0;y4[18]:=0;y4[17]:=1;y4[16]:=1;
y4[15]:=0;y4[14]:=1;y4[13]:=0;y4[12]:=0; y4[11]:=0;y4[10]:=1;y4[9]:=1;y4[8]:=0;
y4[7]:=0;y4[6]:=1;y4[5]:=0;y4[4]:=1; y4[3]:=0;y4[2]:=0;y4[1]:=1;y4[0]:=1;

dodeli(hh1,h1);dodeli(hh2,h2);dodeli(hh3,h3); dodeli(hh4,h4);dodeli(hh5,h5);
m:=duz div 512;
    reset(g);
{Processing}
i:=0;
while i<=m do
begin
    for j:=0 to 15 do
    begin
      for l:=0 to 31 do
             read(g,x[j,l]);
             end;
         for j:=16 to 79 do
       begin
        for l:=0 to 31 do
          p[l]:=(((x[j-3,l] xor x[j-8,l]) xor x[j-14,l]) xor        x[j-16,l]);
             l:=1;
           rot(p,l);
               for l:=0 to 31 do x[j,l]:=p[l];
             end;
          i:=i+1;
         end;
{initialize working variables}
dodeli(aa,hh1);dodeli(bb,hh2);dodeli(cc,hh3); dodeli(dd,hh4);dodeli(ee,hh5);
for j:=0 to 19 do
begin
 dodeli(pp,aa); dodeli(ss,bb);
dodeli(nn,cc); dodeli(mm,dd);
for l:=0 to 31 do qq[l]:=x[j,l];
   fi(bb,cc,dd,rr);
   rot(aa,5);
 saberi(aa,rr,pom);
 saberi(pom,ee,pom);
 saberi(pom,qq,pom);
 saberi(pom,y1,pom);
 for l:=0 to 31 do tt[l]:=pom[l];
 dodeli(aa,tt);dodeli(bb,pp);
   rot(ss,30);
    dodeli(cc,ss);
     dodeli(dd,nn);dodeli(ee,mm);
end; writeln;
for j:=20 to 39 do
begin
 dodeli(pp,aa);dodeli(ss,bb);
  dodeli(nn,cc);
```

```
 dodeli(mm,dd);
  for l:=0 to 31 do qq[l]:=x[j,l];
 hi(bb,cc,dd,rr);
 rot(aa,5);
 saberi(aa,rr,pom);
 saberi(pom,ee,pom);
 saberi(pom,qq,pom);
 saberi(pom,y2,pom);
 for l:=0 to 31 do tt[l]:=pom[l];
 dodeli(aa,tt);
 dodeli(bb,pp);rot(ss,30);dodeli(cc,ss);dodeli(dd,nn); dodeli(ee,mm);
end;
  for j:=40 to 59 do
begin
 dodeli(pp,aa);dodeli(ss,bb);
  dodeli(nn,cc); dodeli(mm,dd);
  for l:=0 to 31 do qq[l]:=x[j,l];
 gi(bb,cc,dd,rr);
 rot(aa,5);
 saberi(aa,rr,pom);
 saberi(pom,ee,pom);
 saberi(pom,qq,pom);
 saberi(pom,y3,pom);
 for l:=0 to 31 do tt[l]:=pom[l];
 dodeli(aa,tt);
 dodeli(bb,pp);rot(ss,30);dodeli(cc,ss);dodeli(dd,nn); dodeli(ee,mm);
end;
 for j:=60 to 79 do
begin
 dodeli(pp,aa);dodeli(ss,bb);
  dodeli(nn,cc); dodeli(mm,dd);
  for l:=0 to 31 do qq[l]:=x[j,l];
 hi(bb,cc,dd,rr);
 rot(aa,5);
 saberi(aa,rr,pom);
 saberi(pom,ee,pom);
 saberi(pom,qq,pom);
 saberi(pom,y4,pom);
 for l:=0 to 31 do tt[l]:=pom[l];
 dodeli(aa,tt);
 dodeli(bb,pp);rot(ss,30);dodeli(cc,ss);dodeli(dd,nn); dodeli(ee,mm); end;
saberi(hh1,aa,pom);
 for l:=0 to 31 do hh1[l]:=pom[l] ;
 saberi(hh2,bb,pom);
  for l:=0 to 31 do hh2[l]:=pom[l] ;
  saberi(hh3,cc,pom);
  for l:=0 to 31 do hh3[l]:=pom[l] ;
   saberi(hh4,dd,pom);
  for l:=0 to 31 do hh4[l]:=pom[l] ;
   saberi(hh5,ee,pom);
  for l:=0 to 31 do hh5[l]:=pom[l] ;
```

```
 writeln('Binary Hash value:');
 writeln;
 for l:=0 to 31 do write(hh1[l]);
 for l:=0 to 31 do write(hh2[l]);
  for l:=0 to 31 do write(hh3[l]);
  for l:=0 to 31 do write(hh4[l]);
  for l:=0 to 31 do write(hh5[l]);
  writeln;
  assign(f,'hash.dat');
  rewrite(f);
  writeln('hex hash value:');   writeln;
  for l:=31 downto 0 do write(f,hh5[l]);
  for l:=31 downto 0 do write(f,hh4[l]);
  for l:=31 downto 0 do write(f,hh3[l]);
  for l:=31 downto 0 do write(f,hh2[l]);
  for l:=31 downto 0 do write(f,hh1[l]);
  for l:=0 to 7 do
  begin
     i1:=hh1[4*l];i2:=hh1[4*l+1];i3:=hh1[4*l+2]; i4:=hh1[4*l+3];
 ses(i1,i2,i3,i4);
 end;
  for l:=0 to 7 do
  begin
     i1:=hh2[4*l];i2:=hh2[4*l+1];i3:=hh2[4*l+2]; i4:=hh2[4*l+3];
 ses(i1,i2,i3,i4);
 end;
 for l:=0 to 7 do
  begin
     i1:=hh3[4*l];i2:=hh3[4*l+1];i3:=hh3[4*l+2]; i4:=hh3[4*l+3];
 ses(i1,i2,i3,i4);
 end;
 for l:=0 to 7 do
  begin
     i1:=hh4[4*l];i2:=hh4[4*l+1];i3:=hh4[4*l+2]; i4:=hh4[4*l+3];
 ses(i1,i2,i3,i4);
 end;
 for l:=0 to 7 do
  begin
i1:=hh5[4*l];i2:=hh5[4*l+1];i3:=hh5[4*l+2]; i4:=hh5[4*l+3];
 ses(i1,i2,i3,i4);
 end; readln; readln;
end.
```

## 3.3. EXAMPLES OF HASH VALUES

The result of this function is the 160 series of zeros and ones whose order depends on the message.

Examle 1: Using this software, we will determine the hash value of the message: *Advanced Computing: An International Journal (ACIJ)*

Output to the screen:

Input message:Advanced Computing: An International Journal (ACIJ)

Binary Hash value:

10111011100000001111001001100000001111011000001001010011111001001111000011011100
00110100111111010111101010100100110110100101000101000101001000110111100110000101

hex hash value:

bb80f2603d8253e4f0dc34fd7aa4da5145237985

Example 2. If we left out (:) in message: *Advanced Computing: An International Journal (ACIJ)*

, we get output to the screen:

Input message:Advanced Computing An International Journal (ACIJ)

Binary Hash value:

00100010100000001110101011100111011110011011011110000101110110000010111101000000
01000111011000011101100100110001110111010001010100010100101100011100001010010010

hex hash value:

2280eae779b785d82f404761d931dd1514b1c292

The omission of a single-letter hash value has undergone drastic changes. Undermined the integrity of the message.

## 4. HOW DIGITAL SIGNATURE LOOK IN REALITY

In this paragraph, we will follow the steps of a message signing by the own software. It can be found in [4].

The first step of a scheme is to detect two large (probably) prime numbers p and q, of approximately the same number of digits. In this paper, we choose two 512-bit numbers that we got by using our software realization of the Miler-Rabin algorithm.

Detected (probably) prime numbers are:

**p:**

100000000000000000000000000000000000000000000000000000000000000000000000000000000000000000000
000000000000000000000000000000000000000000000000000000000000000000000000000000000000000000000
000000000000000000000000000000000000000000000000000000000000000000000010000000000000000000000
000000000000000000000000000000000000000000000000000000000000000000000000000000000000000000000
000000000000100000000000000000000000000100000000000000000000000000000000000000000000000001000
00000000000000000000000000000000000010110011101

**q:**

100000000000000000000000000000000000000000000000000000000000000000000000000000000000000000000
000000000000000000000000000000000000000000000000000000000000000000000000000000000000000000000
000000000000000000000000000000000000000000000000000000000000000000000010000000000000000000000
000000000000000000000000000000000000000000000000000000000000000000000000000000000000000000000
000000000000100000000000000000000000000100000000000000000000000000000000000000000000000000000
00000000000000000000000000000000000011110000011

Using our software from [3], we compute n= p*q as well as $\phi$ = (p-1)*(q-1)

**n=pq:**
100000000000000000000000000000000000000000000000000000000000000000000000000000000000
000000000000000000000000000000000000000000000000000000000000000000000000000000000000
000000000000000000000000000000000000000000000000000000000000001000000000000000000000
000000000000000000000000000000000000000000000000000000000000000000000000000000000000
000000000010000000000000000000000001000000000000000000000000000000000000000000001000
000000000000000000000000000000001101001000010000000000000000000000000000000000000000
000000000000000000000000000000000000000000000000000000000000000000000000100000000000
000000000000100000000000000000000000000000000000000000000001000000000000000000000000
000000000011010010000100000000000000000000000100000000000000000000000100000000000000
000000010000000000000000000000000010000000000110100100000000000000011010010000000000
0000000000000000000000000000111100000110000000000000000000000001010100010101001010111.

ϕ =(p-1)(q-1):
100000000000000000000000000000000000000000000000000000000000000000000000000000000000
000000000000000000000000000000000000000000000000000000000000000000000000000000000000
000000000000000000000000000000000000000000000000000000000000001000000000000000000000
000000000000000000000000000000000000000000000000000000000000000000000000000000000000
000000000010000000000000000000000001000000000000000000000000000000000000000000001000
000000000000000000000000000000001101000111101000000000000000000000000000000000000000
000000000000000000000000000000000000000000000000000000000000000000000000100000000000
000000000000010000000000000000000000000000000000000000000000100000000000000000000000
000000000011010001111010000000000000000000000100000000000000000000000100000000000000
000000010000000000000000000000000010000000000110100011100000000000110100011100000000
0000000000000000000000000000111110000010000000000000000000000001010100001110100111000.

Then, we choose the public key, let’s assume e: 111, and using the same software we solve the equation e*d≡ 1 (mod ϕ), or cryptographically said, we compute the private key[4][6][7].

d:
100100100100100100100100100100100100100100100100100100100100100100100100100100100100
100100100100100100100100100100100100100100100100100100100100100100100100100100100100
100100100100100100100100100100100100100100100100100100100100101001001001001001001001
001001001001001001001001001001001001001001001001001001001001001001001001001001001001
001001001010010010010010010010011011011011011011011011011011011011011011011011100100
100100100100100100100100100100100110011100100001001001001001001001001001001001001001
001001001001001001001001001001001001001001001001001001001001001001001001001101101101
101101101101110010010010010010010010010010010010010010010010010110110110110110110110
110110110111110010110101001001001001001001001010010010010010010010010110110110110110
110111000000000000000000000000001001001001011000001000100100100100110011100100000000
000000000000000000000000000000100010010100100100100100100100100111100101000101111111.

Let “Elektrotehnicki fakultet u Beogradu” be the message we should sign. Its hash value is:

m:
00111111000111001010001001000111101110111010001100111111010000111100111101100011000011000110
111010010010000100010100001001101110010010011100000101101000011011

The digital signature of a message m hash value is s= $m^d$ mod n.

s:
101110110001100000000111000100011001011111110100111001101010011001010000101110010100000110010
110101110110001110000101111111111000000100100011000001010001110111111000010010010000011100001101
000100011100011110000101101010100110100100111111000111000110000001110011110101010100001111110
0101101111111010001110011001011001001100011001110000001011110011000010111100000010100101101110
001000001100010100001100010110001101101110001101110110110100111110010100001001011001110011001
010010100100011010001000011111101011001110011010100110111000011001111111011101001011101000101
101011111011000010111000010100010100010111011100010011010001100011000011000101000010001011010

10111000100010011000101110111111011110010000010000101110000011110110110000000011111000011100111101011101111110111110110000001111110001001101000010011111111011101010101111001011011001111111101111101101111101101100001101100110011110110001110111011011110000101101100001000000001111010101010000110101011001011100100001010101010100110011010111110110101101110111110110000110101001.

If we check it, we get $m' = s^e \bmod n$.

m':

0011111100011100101000100100011110111011101000110011111101000011110011111011000110000110001101110100100100001000101000010011011100100100111000001011010000110110.

By this, we are sure that using the previous operation, we really get the same value (m=m’). it means that the data integrity is preserved and that the owner of a private key is the one who signed the message.

# 5. FUTURE WORK

In the arguments for and against in a trial of strength of ECC **(Elliptic Curve Cryptography)** and RSA, the simple fact that they are performed by the same tools made for operations with large numbers, is usually overlooked. Mathematical bases of RSA and ECC are completely different [2] [8], but they need the same operations: addition, subtraction, multiplication, division, finding the remainder, calculating d from the equation $e*d \equiv 1 \pmod{p}$ for fixed values of e and p, SHA-1 and more other joint auxiliary operations needed for the realization of a digital signature in both schemes. Therefore, ECC is our next goal-because we have the tools.

# 6. CONCLUSION

We believe that each country must stimulate young people’s interest in cryptography, because we doubt that our secret data can be protected using someone else’s software.

Of course, it is very difficult to develop our own protection mechanisms, but we think it is far better to protect data using our own mechanisms first, and then, thus modified, leave them to someone else’s software, than to allow the original data be protected by somebody else’s mechanisms, which is a logical nonsense.

That is the reason why we always insist on more our own softwares and a greater interest in cryptography, which seems itself (in case it wasn’t brought closer to a reader) pretty cryptic and bouncing[5]. So, this work is primarily addressed to young researches as an incentive to try to develop their own tools for data protection. Those tools do not have to be flawless, they may be far below the level of the tools found on the market. However, they should be good enough for the beginning of a hard work that would lead researches to some great commercial solutions.